# Coherent Controllable Transport of a Surface Plasmon Coupled to Plasmonic Waveguide with a Metal Nano Particle-Semiconductor Quantum Dot Hybrid System

Myong-Chol Ko,[1] Nam-Chol Kim,[1,2,*] Zhong-Hua Hao,[2] Li Zhou,[2] Jian-Bo Li,[3] and

Qu-Quan Wang[2,4 *]

[1]Department of Physics, **Kim Il Sung** University, Pyongyang, DPR of Korea
[2]School of Physics and Technology, Wuhan University, Wuhan 430072, China
[3]Institute of Mathematics and Physics, Central South University of Forestry and Technology, Changsha 410004, China
[4]The Institute for Advanced Studies, Wuhan University, Wuhan 430072, China
*ryongnam10@yahoo.com , qqwang@whu.edu.cn

**Abstract:** By using the real-space method, switching of a single plasmon interacting with a hybrid nanosystem composed of a semiconductor quantum dot (SQD) and a metallic nanoparticle (MNP) coupled to one-dimensional surface plasmonic waveguide is investigated theoretically. We discussed that the dipole coupling between an exciton and a localized surface plasmon results in the formation of a hybrid exciton and the transmission and reflection of the propagating single plasmon could be controlled by changing the interparticle distance between the SQD and the MNP and the size of the nanoparticles. The controllable transport of the propagating single surface plasmon by such a nanosystem discussed here could find the significant potential in the design of next-generation quantum devices such as plasmonic switch, single photon transistor and nanolaser and quantum information.

[*] Electronic mail: ryongnam10@yahoo.com,  qqwang@whu.edu.cn

## 1. Introduction

The light interacted with matter has always become a hot research topic in nanotechnology for some fundamental investigations of photon-atom interaction and for its significant potential in active device fabrication for all-optical switching or information processing, and its most elementary level is the interaction between a single photon and a single quantum emitter [1, 2]. Recently, the coherent controlling of a single surface plasmon (that is, single photon) transport is a central topic in quantum information and nanodevices, because it is the most efficient means to send information to and obtain information from nanoscale, and theoretical idea of a single photon transistor for switching light by light on the nanoscale has also reported [3].

The photon scattering in different quantum systems has been studied in many theoretical [4-10] and experimental [11-15] works to accomplish the coherent controlling of the propagating single photon. In recent researches, the metallic nanowire is used as an intermediary in the transport of the single photon because it is useful to guide and manipulate light on nanoscale. It is useful to resort to real space approach for determination of the response to the single injected photon, which is particularly convenient for discussing photon transport from one space-time point to another one and makes no assumptions on temporal behaviors of the constituents of the system. Therefore, a single photon transport based on the real-space method [4, 5] has a period of explosive growth, including the scattering properties of the single photon interacting with quantum emitters which have various structures [4-10, 16-19]. Recently, optical properties of a complex nanosystem which combines a SQD and a MNP, such as a nonlinear Fano effect [19], the creation of controlled Rabi oscillations [20], plasmonic meta-resonances [21], tunable ultrafast nanoswitches [22], modified resonance fluorescence and photon statistics [23, 24], and intrinsic optical bistability [25] have been also investigated widely. Especially, it has been reported that the plasmon-exciton interaction leads to the formation of a hybrid exciton for the energy absorption of the hybrid system and the control of energy dissipation [20, 23]. However, the case where an incident single plasmon interacts with a hybrid system composed of a SQD and a MNP has not been investigated yet. In this paper, we investigate theoretically the controllable transport of a single plasmon interacting with a hybrid nanosystem composed of a semiconductor quantum dot and a metallic nanoparticle coupled to 1D surface plasmonic waveguide.

## 2. Theoretical Model and Dynamics Equations

The schematic diagram of the system considered in this paper is exhibited in Fig. 1, where the hybrid system consists of a spherical MNP of radius $a$ and a spherical SQD with radius $r$ in the presence of polarized classic optical induced field with the frequency $\omega_c$, $E = E_0 e^{-i\omega_c t}/2$. The center-to-center distance between the MNP and the SQD is denoted as $R$. For the description of the MNP, we use classical electrodynamics and the quasi-static approach. For the spherical SQD, we employ the density matrix formalism and the following model with two levels.

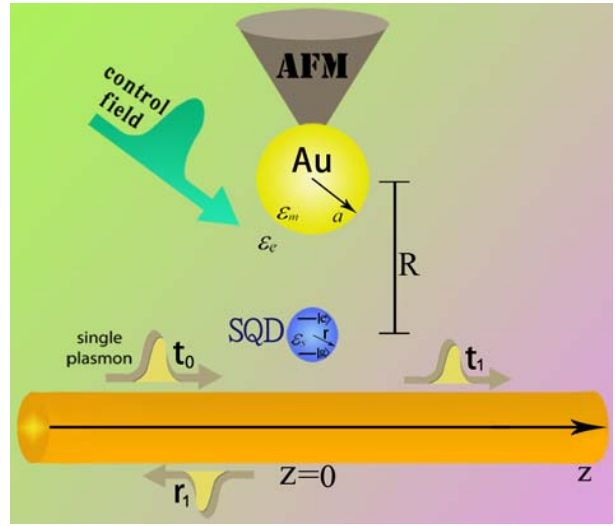

**Fig. 1** (Color online). The schematic diagram of the hybrid nanosystem coupled to one Dimensional waveguide. $t_1$ and $r_1$ are the transmission and reflection amplitudes at the place $z$, respectively, where $\varepsilon_e$, $\varepsilon_m$ and $\varepsilon_s$ are the dielectric constants of the background medium, the MNP and the SQD and $a$ is a radius of the spherical MNP and $r$ is a radius of the spherical SQD, $R$ is the interparticle distance between the MNP and the SQD.

### 2.1. Hybrid Exciton with the Exciton-Plasmon Coupling

Firstly, we discuss that the exciton-plasmon dipole coupling effect results in the formation of an excitonic system with a natural frequency different from the resonant frequency of the single SQD. The Hamiltonian of the SQD in the hybrid system consisting of a single plasmon and a two-level SQD shown in Fig. 1 can be written as follows:

$$\hat{H}_{SQD} = \hbar\omega_0 b^+ b - \mu_{ge} E_{SQD} b - \mu_{ge} E^*_{SQD} b^+, \qquad (1)$$

where $\omega_0$ is the frequency associated with the transition $|g\rangle - |e\rangle$, $b^+(b)$ is the creation(annihilation) operator for the ground state and the exciton state of the SQD and $\mu_{ge}$ is the dipole moment associated with the transition between the ground state and the exciton state in the SQD. $E_{SQD}$ is the total electric field experienced by the SQD and consists of the classic optical induced field, $E$, and the internal field produced by the polarization of the MNP, $P_{MNP}$. In the dipole limit, $E_{SQD}$ can be written as $E_{SQD} = E + (s_\alpha P_{MNP} / 4\pi\varepsilon_e \varepsilon_{eff} R^3)$, where $\varepsilon_{eff} = (2\varepsilon_e + \varepsilon_s)/3\varepsilon_e$, and $\varepsilon_e$ and $\varepsilon_s$ are the dielectric constants of the environment medium and the SQD, respectively. $s_\alpha = 2(-1)$, when the classic optical induced field polarization is parallel (perpendicular) to the major axis of the hybrid system. The dipole $P_{MNP}$ comes from the charge induced on the surface of the MNP. It depends on the total electric field which is the superposition of classic optical induced field and the dipole field due to the SQD, $P_{MNP} = (4\pi\varepsilon_e)\gamma a^3 [E + (s_\alpha P_{SQD} / 4\pi\varepsilon_e \varepsilon_{eff} R^3)]$, where $\gamma = (\varepsilon_m(\omega_c) - \varepsilon_e)/[2\varepsilon_e + \varepsilon_m(\omega_c)]$ and $\varepsilon_m(\omega_c)$ is the dielectric constant of metal. We employ the density matrix $\rho$ to calculate the polarization of the SQD. We label the ground state (no exciton) of the SQD as $|g\rangle$ and the excited state (one exciton) as $|e\rangle$. We then have $P_{SQD} = \mu(\rho_{21} + \rho_{12})$. Considering above relations, $E_{SQD}$ can be written as follows:

$$E_{SQD} = E\left[1 + \frac{s_\alpha \gamma a^3}{\varepsilon_{eff} R^3}\right] + \frac{s_\alpha^2 \gamma a^3 P_{SQD}}{4\pi\varepsilon_e \varepsilon_{eff}^2 R^6} \qquad (2)$$

Factoring out the high-frequency time-dependence of the off-diagonal terms of the density matrix as $\rho_{12} = \tilde{\rho}_{12} e^{i\omega_c t}$, we obtained the following equations to determine the density matrix of the SQD coupled to the MNP via Coulomb interaction in the steady state limit with $\tilde{\rho}_{12} = A + iB$.

$$-\frac{A}{T_{20}} + (\omega - \omega_0)B + (\Omega_I + \Gamma_{FRET} A + \omega_{shift} B)\Delta = 0, \qquad (3.\ a)$$

$$-\frac{B}{T_{20}} - (\omega - \omega_0)A - (\Omega_R + \omega_{shift}A - \Gamma_{FRET}B)\Delta = 0, \quad (3.\,b)$$

$$\frac{(1-\Delta)}{\tau_0} - 4A\Omega_I + 4B\Omega_R - 4\Gamma_{FRET}(A^2 + B^2) = 0, \quad (3.\,c)$$

where $\Delta = \rho_{11} - \rho_{22}$, $\Omega_R = \text{Re}[\Omega_{eff}]$, $\Omega_I = \text{Im}[\Omega_{eff}]$. Here, we defined the following parameters as $\Omega_{eff} = \Omega_0 \left[1 + \frac{s_\alpha \gamma}{\varepsilon_{eff}} \left(\frac{a}{R}\right)^3\right]$, $\eta = \frac{s_\alpha^2 \gamma a^3 \mu^2}{4\pi \varepsilon_e \hbar \varepsilon_{eff}^2 R^6}$, where $\Omega_0 = \frac{\mu E_0}{2\hbar}$ refers to the Rabi frequency of the SQD in the absence of the MNP (no plasmonic or coherent effect). $\Omega_{eff}$ represents the Rabi frequency of the SQD normalized by the plasmonic effects. On the other hand, the imaginary part of $\eta$ describes the FRET (Förster resonance energy transfer) rate, $\Gamma_{FRET}$, from the SQD to the MNP. The real part of $\eta$ shows the plasmonic shift of the exciton transition (the $|g\rangle - |e\rangle$ transition), $\omega_{shift}$. In Eq. (3.c), the relaxation time $\tau_0$ represents a contribution from nonradiative decay to dark states. The decay rate from the exciton state to the ground state, $\Gamma_{eg}$, is represented with the life time of the exciton state, $T_{20}$, as $\Gamma_{eg} = \Gamma_{ge}^* = \rho_{12}/T_{20}, T_{20} = 0.3(\text{ns})$. For a weak external field, we have the following steady state solution in the analytical form $\tilde{\rho}_{12} = -\Omega_{eff} / \{[(\omega_c - \omega_0) + \omega_{shift}] - i(\Gamma_{eg} + \Gamma_{FRET})\}$.

## 2.2. Controllable Transport of the Propagating Single Plasmon with the MNP-SQD Hybrid System

Now, we consider the transport properties of a propagating single plasmon interacting with the MNP-SQD hybrid system. We consider the MNP-SQD hybrid system as a new exciton with the transition frequency of which is a very frequency of the hybrid exciton of the MNP-SQD hybrid system and its transition frequency is mediated by environment field fluctuation. Under the rotating wave approximation, the Hamiltonian of the system in real space is given by

$$\begin{aligned}H/\hbar &= (\omega_2 - i\Gamma/2)\sigma_{22} + \omega_1 \sigma_{11} + iv_g \int_{-\infty}^{\infty} dz[a_l^+(z)\partial_z a_l(z) - a_r^+(z)\partial_z a_r(z)] \\ &\quad + g\{[a_r^+(z) + a_l^+(z)]\sigma_{12} + [a_r(z) + a_l(z)]\sigma_{21}\}\end{aligned} \quad (4)$$

Here, $\omega_1$ and $\omega_2$ are the eigenfrequencies of the state $|1\rangle$ and $|2\rangle$ of the hybrid exciton, respectively, $\omega_{sp}$ is the frequency of the surface plasmon with wavevector $k$ ($\omega_{sp} = v_g |k|$). $\sigma_{12} = |1\rangle\langle 2|\ \sigma_{21} = (|2\rangle\langle 1|)$ is the lowing (raising) operators of the hybrid exciton, $a_r^+(z)(a_l^+(z))$ is the bosonic operator creating a right-going (left-going) plasmon at position $z$ of the hybrid exciton. $v_g$ is the group velocity corresponding to $\omega_{sp}$, the non-Hermitian term in $H$ describes the decay of state $|2\rangle$ at a rate $\Gamma = \Gamma_{eg} + \Gamma_{FRET}$ into all other possible channels. $g = (2\pi\hbar/\omega_{sp})^{1/2}\Omega \mathbf{D} \cdot \mathbf{e}_k$ is the coupling constant of the hybrid exciton with a single plasmon, $\Omega$ is the resonance energy of the hybrid exciton, **D** is the dipole moment of the hybrid exciton, $\mathbf{e}_k$ is the polarization unit vector of the surface plasmon [4, 5]. The Hamiltonian includes three parts. The first term describes the hybrid exciton, the second term describes propagating single plasmons which run in both directions and the third term describes the interaction between the hybrid exciton and the single propagating plasmon. Assuming that a single plasmon is incoming from the left with energy $E_k = \hbar\omega_{sp}$ then the eigenstate of the system, defined by $H = E_k |\psi_k\rangle$, can be constructed in the form

$$|\psi_k\rangle = \int dz [\phi_{k,r}^+(z) a_r^+(z) + \phi_{k,l}^+(z) a_l^+(z)] |0,1\rangle + e_k |0,2\rangle \qquad (5)$$

where $|0,1\rangle$ denotes the vacuum state with zero plasmon and the hybrid exciton being unexcited, $|0,2\rangle$ denotes the hybrid exciton in the excited state and $e_k$ is the probability amplitude of the hybrid exciton in the excited state. $\phi_{k,r}^+(z)$ ($\phi_{k,l}^+(z)$) is the wavefunction of a right-going (a left-going) plasmon at position $z$.

Now, we can solve the Schrodinger equations by substituting Eq. (5) into Eq. (4). For a single plasmon incident from the left, the mode functions $\phi_{k,r}^+(z)$ and $\phi_{k,l}^+(z)$ can take the forms as followings, $\phi_{k,r}^+(z<0) = e^{ikz}$, $\phi_{k,r}^+(z>0) = te^{ikz}$, $\phi_{k,l}^+(z<0) = re^{-ikz}$, $\phi_{k,l}^+(z>0) = 0$. Here $t$ and $r$ are the transmission and reflection amplitudes at the place $z$, respectively. By taking the boundary conditions of the mode functions into account, we obtain the following equations; $-gr_1 - iJe_k = 0$, $gt_1 - gt_0 + iJe_k = 0$, $g(t_0 + r) + \Delta e_k = 0$, where $\Delta_k = \Omega - \omega_{sp}$, $g^2/v_g \equiv J$ and $\Delta = \Delta_k - i\Gamma/2$ and then we can get the transmission and the reflection amplitudes, respectively, where $t_0 = 1$ and $r_2 = 0$ into account. By evaluating the transmission coefficient $T_1 \equiv |t_1|^2$ and the reflection coefficient $R_1 \equiv |r_1|^2$, we can obtain the transport properties of a single plasmon in the long time limit.

### 3. Theoretical Analysis and Numerical Results

First of all, we investigate the plasmonic effects on the SQD in MNP-SQD hybrid system and the resonant frequency of the hybrid exciton versus the frequency of the classic optical induced field in different inter-particle distances, respectively [Fig. 2], where the imaginary and real part of $\eta$ represent the FRET rate and the shift of the frequency, respectively.

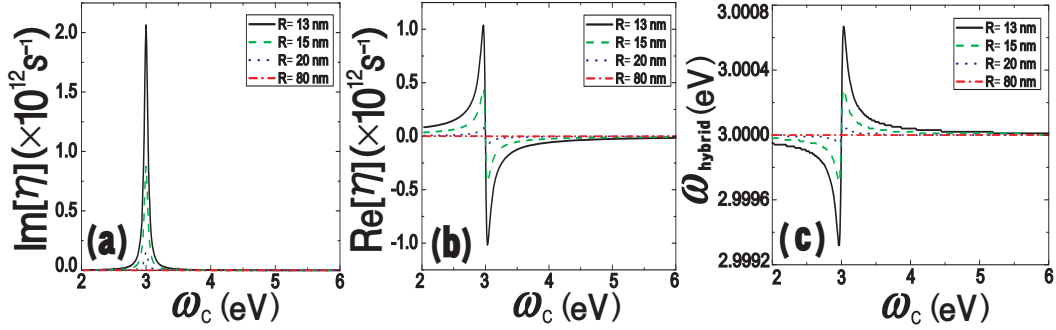

**Fig. 2** (Color online). The plasmonic effect of the Au nanoparticle on the SQD and the resonant frequency of the hybrid exciton versus the frequency of the classic optical induced field when the distances between the MNP and the SQD are $R$=12nm(solid line), $R$=15nm(dashed line), $R$=20nm(dotted line), $R$=80nm(dash-dotted line), respectively. (a) The imaginary part of $\eta$, (b) the real part of $\eta$, (c) the resonant frequency of the hybrid system composed of the Au nanoparticle and the SQD, where $\varepsilon_e = 3$, $\varepsilon_s = 6$, $\omega_0 = 3\,\text{eV}$, $a = 7\,\text{nm}$ and $\mu = 10^{-28}\,\text{C}\cdot\text{m}$.

From Fig. 2, we can see the dipole-dipole interaction between the MNP and SQD is very strong when the frequency of the classic optical induced field is equal to the natural frequency of the SQD. It can also be found that the influence of plasmonic effect of the MNP on the SQD is drastically decreased as the frequency of the classic optical induced field is biased more and more from the natural frequency of the SQD. Fig. 2(a) shows the imaginary part of $\eta$, which represents the FRET rate from the SQD to MNP. We found that the FRET rate is strongly enhanced when $\omega_0 = \omega_c = 3\,\text{eV}$ and the interparticle distance is small. As the distance between the SQD and MNP is decreased, the FRET rate is drastically decreased and it is almost disappeared when $R = 80\,\text{nm}$. In Figs. 2(b) and 2(c), we find that the smaller the interparticle distance becomes, the larger the width of the frequency range of the hybrid exciton becomes, considerably different from the natural frequency of the SQD. When $R = 13\,\text{nm}$, for example, the shift of the frequency of the hybrid exciton from that of the SQD is much larger than the other cases, and when $R = 80\,\text{nm}$, the shift of the frequency of the hybrid exciton disappeared. From the results discussed above, we found that the shifted frequency of the hybrid exciton is very sensitive to the interparticle distance between the SQD and MNP, which suggests one can control the frequency of the hybrid MNP-SQD system by adjusting the interparticle distances effectively.

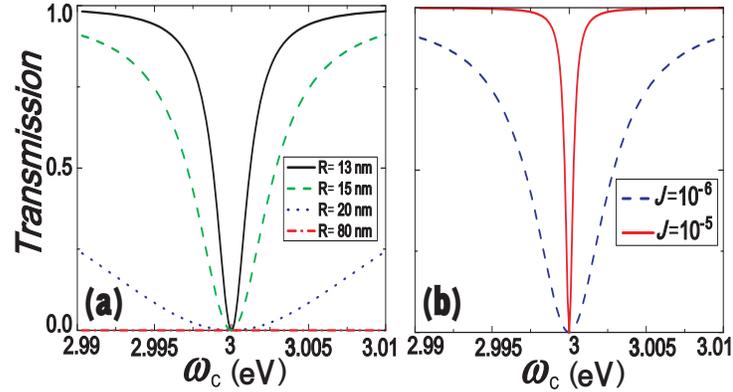

**Fig. 3** (Color online). The transmission spectra of propagating plasmon interacting with the hybrid system versus the frequency of the classic optical induced field, where $\varepsilon_e = 3$, $\varepsilon_s = 6$, $\omega_0 = 3\,\text{eV}$, $\omega_{sp} = 3\,\text{eV}$, $a = 7\,\text{nm}$, $\mu = 10^{-28}\,\text{C}\cdot\text{m}$ : (a) the transmission spectra when $J = 10^{-6}\omega_0$ with different distances between the MNP and the SQD and (b) the transmission spectra with different coupling strength between the hybrid system and one dimensional waveguide when $R = 15\,\text{nm}$.

Figure 3 shows the transmission spectra of the propagating plasmon could be controlled by adjusting the interparticle distance between the SQD and MNP and the coupling strength between the hybrid system and the 1D plasmonic waveguide. As shown in Fig. 3(a), the scattering spectrum shows a complete reflection at $\omega_c = 3\,\text{eV}$ when $R$=13nm. As the interparticle between the SQD and MNP distance increases, the width of scattering spectrum increases, and the complete reflection peak(not shown in Fig. 3(a)) disappears when $R$=80nm. The above result suggests the exciton-plasmon coupling effect could be neglected when the interparticle distances longer than about 80nm for the other parameters set in this paper, which is quite different from the previous results[4-10], where it was found that there will be a single complete reflection peak when the frequency of the propagating plasmon is resonant with that of the bare exciton. However, our result shows that there could not be appeared a complete reflection peak even when $\omega_0 = \omega_{sp}$ because of the plasmonic effect of the MNP on the SQD at the frequencies of the classic optical induced field different from 3eV. Fig. 3(b) shows the transmission spectrum of the propagating plasmon in the different coupling strengths between the hybrid system and the 1D waveguide. As shown in Fig. 3(b), the line width becomes narrower as the coupling strength becomes stronger, which suggest one can control the switching of a single plasmon very sensitively by adjusting the coupling strength between the hybrid system and the plasmonic waveguide.

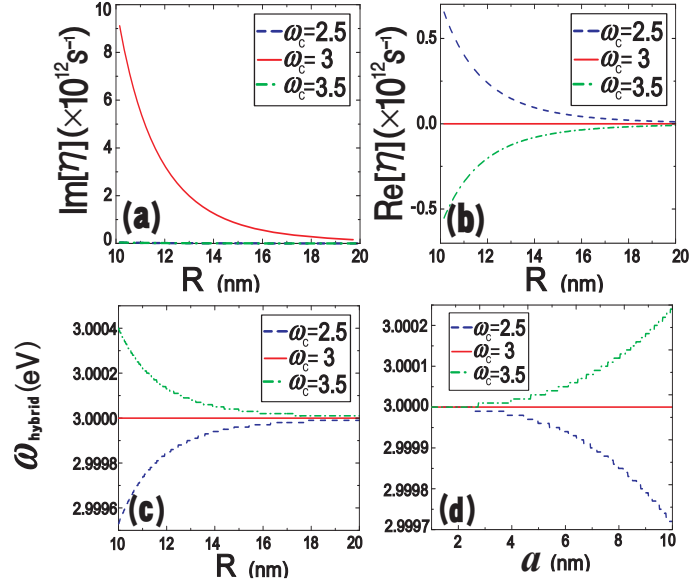

**Fig. 4** (Color online). The influence of the interparticle distance and the size of MNP on the exciton-plasmon coupling when the frequencies of the classic optical induced field are $2.5\,\text{eV}$ ( dashed line), $3\,\text{eV}$ ( solid line), $3.5\,\text{eV}$ ( dash-dotted line), respectively. (a) The imaginary part of $\eta$ versus interparticle distance, (b) the real part of $\eta$ versus interparticle distance, (c) the frequency of the hybrid exciton versus interparticle distance, where $\varepsilon_e = 3$, $\varepsilon_s = 6$, $\omega_0 = 3\,\text{eV}$, $a = 7\,\text{nm}$, $\mu = 10^{-28}\,\text{C}\cdot\text{m}$, and (d) the frequency of the hybrid exciton versus the size of Au MNP when *R*=13.

Now, we consider the influence of the interparticle distance between the SQD and MNP and the size of the MNP on the exciton-plasmon coupling with different frequencies of the classic optical induced field [Fig. 4]. In Fig. 4(a), we find there appears a positive branch only when $\omega_c = 3\,\text{eV}$, and the imaginary part of $\eta$ decreases rapidly as the interparticle distance between the SQD and MNP increases, which implies the FRET rate depends on the interparticle distance sensitively. It can also be found that there does not exist the FRET rate from the SQD to MNP when $\omega_c = 2.5\,\text{eV}$ and $\omega_c = 3.5\,\text{eV}$ even though the frequency of the incident single plasmon is resonant with the natural frequency of the SQD, which could be used as a way to decrease the dissipation of the hybrid system. As shown in Figs 4(b) and 4(c), when $\omega_c = \omega_0 = 3\,\text{eV}$, that is, the frequency of the classic optical induced field is resonant with the frequency of the SQD, there is no the frequency shift between the natural frequency of the SQD and the frequency of the hybrid exciton. However, in the case of $\omega_c = 2.5\,\text{eV}$ ($\omega_c = 3.5\,\text{eV}$), the shift of the frequency has a positive (negative) value, resulting in the red(blue) detuning of the frequency of the hybrid exciton, the value of which decreases rapidly as the interparticle distance increases. On the other hand, Fig. 4(d) shows the influence of the size of the MNP on the transition frequency of the hybrid exciton, from which one can find that the frequency of the hybrid exciton could be controlled with the size of the MNP to some extent. We can also find that the increasing the size of the nanoparticles is effectively relevant to the decreasing the interparticle distance.

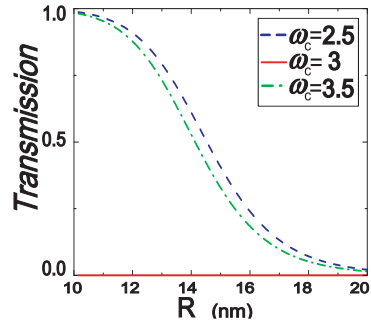

**Fig. 5** (Color online). The transmission spectrum of the propagating plasmon versus the interparticle distance when the frequencies of the classic optical induced field are $2.5\,\text{eV}$ (dashed line), $3\,\text{eV}$ (solid line), $3.5\,\text{eV}$ (dash-dotted line), respectively. Here we set $\varepsilon_e = 3$, $\varepsilon_s = 6$, $\omega_0 = 3\,\text{eV}$, $\omega_{sp} = 3\,\text{eV}$, $a = 7\,\text{nm}$, $\mu = 10^{-28}\,\text{C}\cdot\text{m}$.

On the basis of the above results, we investigate the transmission properties of the propagating plasmon versus the interparticle distance when the frequencies of the classic optical induced field are $\omega_c = 2.5, 3, 3.5\,(\text{eV})$, respectively. From Fig. 5, we could find the transport properties of the propagating plasmon could be controllable by using the inter-particle distance and the frequency of the classic optical induce field. As we can see from Fig. 5, the transmission coefficient of the propagating plasmon is not zero even when the frequency of the propagating plasmon is resonant with the frequency of the hybrid system, which is also similar to that shown in Fig. 3. The transmission coefficient of the propagating plasmon ranges from 1 to 0, especially approaches asymptotically to 0 as the interparticle distance increases in the range where it is larger than about 20nm when $\omega_c = 2.5\text{eV}$ and $\omega_c = 3.5\text{eV}$. However, the transmission coefficient is always 1 irrelevant to the interparticle distance, when $\omega_c = 3\text{eV}$. The results shown in Fig. 5 suggest switching of an incident single plasmon could be implemented by adjusting the frequency of the classic optical induce field and the interparticle distance.

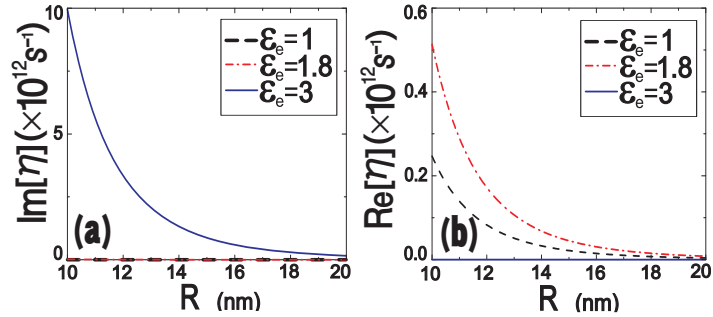

**Fig. 6** (Color online). The influence of the interparticle distance on the exciton-plasmon coupling when the dielectric constants of the background are $\varepsilon_e = 3\text{eV}$ (solid line), $\varepsilon_e = 1.8$ (dash-dotted line) and $\varepsilon_e = 1$ (dashed line), respectively. (a) The imaginary part of $\eta$ and (b) the real part of $\eta$. Here, we set $\omega_c = 3\text{eV}$, $\omega_{sp} = 3\text{eV}$, $\omega_0 = 3\text{eV}$, $a = 7\text{nm}$, $\mu = 10^{-28}\text{C}\cdot\text{m}$ and $\varepsilon_s = 6$.

Next, we consider the influence of the interparticle distance on the exciton-plasmon coupling effect in the hybrid nano-structure when the dielectric constants of the background are $\varepsilon_e = 1$, $\varepsilon_e = 1.8$, and $\varepsilon_e = 3$, respectively. Fig. 6(a) illustrates the FRET rate exists when the dielectric constant of the background, $\varepsilon_e$, is equal to 3, and it decreases rapidly as the interparticle distance increases. When $\varepsilon_e = 1$ and $\varepsilon_e = 1.8$, the FRET rate does not exist for any interparticle distance, which is a way to decrease the dissipation in the hybrid nanosystem. As shown in Fig 6(b), there is no the frequency shift for any interparticle distance when $\varepsilon_e = 3$. In the case of $\varepsilon_e = 1.8$, the shift of the frequency is larger and approaches more rapidly than in the case of $\varepsilon_e = 1$. When $\varepsilon_e = 1$ and $\varepsilon_e = 1.8$, the frequency of the hybrid exciton is red-shifted.

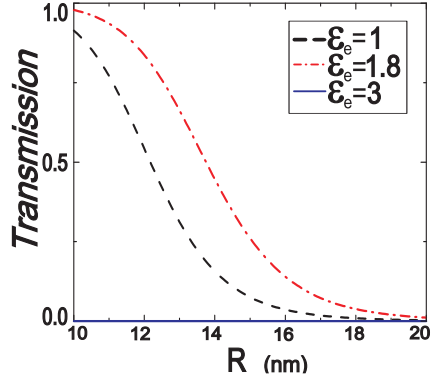

**Fig. 7** (Color online). The transmission of the propagating plasmon versus the inter-particle distance when the dielectric constants of the background are $\varepsilon_e = 3\,\text{eV}$ (solid line), $\varepsilon_e = 1.8$ (dash-dotted line) and $\varepsilon_e = 1$ (dashed line), respectively. Here, we set $\omega_c = 3\,\text{eV}$, $\omega_{sp} = 3\,\text{eV}$, $\omega_0 = 3\,\text{eV}$, $a = 7\,\text{nm}$, $\mu = 10^{-28}\,\text{C}\cdot\text{m}$ and $\varepsilon_s = 6$.

Based on the above results illustrated in Fig. 6, we investigate the transmission spectra of the propagating plasmon versus the interparticle distance when the frequencies of the classic optical induced field are 1, 1.8 and 3, respectively. From Fig. 7, we could find the transport properties of the propagating plasmon interacting with the hybrid system could be controlled by adjusting the interparticle distance between the SQD and the MNP, and the frequency of the classic optical induce field. In Fig. 7, we found the transmission of the propagating plasmon is not zero even when the frequency of the propagating plasmon is resonant with the frequency of the hybrid system. For example, when the frequencies of the classic optical induced field are 1 or 1.8, the transmission coefficient of the propagating plasmon decreases rapidly from 1 to 0, as the interparticle distance increases about to 20nm. Meanwhile, the transmission is equal to 1 for any interparticle distances when $\varepsilon_e = 3$, which implies the MNP-SQD hybrid system could also be utilized in the design of nanomirrors.

## 4. Conclusions

In summary, we studied theoretically the switching of a single plasmon interacting with a hybrid system composed of a semiconductor quantum dot (SQD) and a metal nano-particle (MNP) coupled to 1D surface plasmonic waveguide via the real-space approach. We considered that the synthesizing the MNP and the SQD results in forming of a hybrid exciton, the frequency of which is different from the natural frequency of a bare exciton. The shift of the transition frequency of the hybrid exciton is related to the real part of $\eta$, which can be controlled by adjusting parameters, such as the classic optical induced field, the dielectric constant of the background, the interparticle distant, and so on. Our results show that switching of a single plasmon could be controlled by adjusting the classic optical induced field, the dielectric constant of the background, the inter-particle distant whether the incident frequency of surface plasmon is equal to the natural frequency of the SQD or not. The results illustrated in this paper could be exploited to probe the separation of the nanoparticles and could find the applications in the design of next-generation quantum devices and quantum information, such as quantum switches and nanomirrors.

**Acknowledgments.** This work was supported by the National Program of DPR of Korea (Grant No. 131-00). This work was also supported by the National Program on Key Science Research of China (2011CB922201) and the NSFC (11174229, 11204221, 11374236, 11404410, and 11174372).